\documentstyle[11pt,aaspp4,psfig]{article}
\def\bfx{{\bf x}}
\def\bft{{\vec\gamma}}
\def\nLC{{n^{\rm LC}}}
\def\barnLC{{\bar n^{\rm LC}}}
\def\rmax{{r_{\rm max}}}
\def\rmin{{r_{\rm min}}}
\def\zmax{{z_{\rm max}}}
\def\zmin{{z_{\rm min}}}
\def\VLC{{V^{\rm LC}}}
\def\bfR{{\bf R}}
\def\bfk{{\bf k}}

\def\pp{\par\parshape 2 0truecm 15.5truecm 1truecm 14.5truecm\noindent}
\newcommand{\simgt}{\lower.5ex\hbox{$\; \buildrel > \over \sim \;$}}
\newcommand{\simlt}{\lower.5ex\hbox{$\; \buildrel < \over \sim \;$}}
\newcommand{\himpc}{{\hbox {$h^{-1}$}{\rm Mpc}} }
\begin{document}

\begin{minipage}[c]{4cm}
HUPD-9822\\
RESCEU-49/98\\
UTAP-301/98\\
\end{minipage}\\

\title{ 
{\bf\Large Two-point correlation function of high-redshift objects:}\\
~\\
{\bf\Large  an explicit formulation on a light-cone hypersurface}
}

\bigskip

\author{Kazuhiro Yamamoto \\ {\it Department of Physics, Hiroshima
    University, Higashi-Hiroshima 739-8526, Japan.} \\ and \\ 
  Yasushi Suto \\ {\it Department of Physics and Research Center for
    the Early Universe (RESCEU) \\ School of Science, University of
    Tokyo, Tokyo 113-0033, Japan.} }

\bigskip

\affil{ e-mail: yamamoto@astro.phys.sci.hiroshima-u.ac.jp, 
suto@phys.s.u-tokyo.ac.jp}

\received{1998 August 31}
\accepted{1998 December 9}

\begin{abstract}
  While all the cosmological observations are carried out on a
  light-cone, the null hypersurface of an observer at $z=0$, the
  clustering statistics has been properly defined only on the
  constant-time hypersurface. We develop a theoretical formulation for
  a two-point correlation function on the light-cone, and derive a
  practical approximate expression relevant to the discussion of
  clustering of high-redshift objects at large separations. As a
  specific example, we present predictions of the two-point
  correlation function for the Durham/AAT, SDSS and 2dF quasar
  catalogues. We also briefly discuss the effects of adopted
  luminosity function, cosmological parameters and bias models on the
  correlation function on the light-cone.
\end{abstract}

\vspace*{1cm}
\centerline{\sl The Astrophysical Journal 517 (1999) May 20 issue, in press.}

\clearpage

\baselineskip=13pt

\section{Introduction}

Improving magnitude-limits of astronomical surveys naturally increases
the fraction, and therefore the weight, of selected objects towards
higher redshifts in the entire sample.  In fact discussion of
clustering of objects at $z= (1\sim 3)$ is becoming fairly common,
including the Lyman-break galaxies (Steidel et~al.  1998; Jing \& Suto
1998), X-ray selected AGNs (Carrera et~al. 1998), the FIRST survey
(Magliocchetti et~al. 1998), and up-coming 2dF (2 degree Field; Boyle
et al. 1998) and SDSS (Sloan Digital Sky Survey) QSO surveys.  The
clustering statistics of such high-z objects provides several
important pieces of cosmological information albeit in a rather
complicated manner such as linear and nonlinear evolution of mass
density fluctuations (Hamilton et~al.  1991; Jain, Mo, \& White 1995;
Peacock \& Dodds 1994,1996), evolution of object-dependent bias (Mo \&
White 1996; Jing 1998; Fang \& Jing 1998), and redshift-space
distortion (Kaiser 1987; Hamilton 1997; Ballinger, Peacock \& Heavens
1996; Matsubara \& Suto 1996).  

Another important but less often discussed point is the light-cone
effect, that is, such cosmological observations are feasible only on
the light-cone hypersurface defined by an observer at $z=0$.  In the
case of an angular two-point correlation function of high-$z$ objects,
the light-cone effect can be included relatively easily (e.g., Lahav
et al. 1997; Yamamoto \& Sugiyama 1998), because all the positions of
objects are projected on a two-dimensional sphere on the sky. In
discussing the spatial two-point correlation function of these
objects, however, the light-cone effect hampers any attempt to
distinguish the scale-dependence of clustering in the survey volume
from the intrinsic redshift evolution (e.g., change of the mean number
density of the objects considered).

Some aspects of the light-cone effect have been already discussed by
Matarrese et~al. (1997), Matsubara, Suto \& Szapudi (1997), and
Nakamura, Matsubara \& Suto (1998). In those papers the light-cone
effect is taken into account by integrating over the line-of sight
convolved with the selection function. Matarrese et~al. (1997), for
instance, adopted an expression:
\begin{equation}
\xi_{\rm obs}(r) = \frac{\displaystyle \int\int dz_1 dz_2 
{dN \over dz}(z_1){dN \over dz}(z_2) 
b_{\rm eff}(z_1) b_{\rm eff}(z_2) \xi\left(r, {z_1+z_2 \over2}\right)}
{\displaystyle \left[\int dz {dN \over dz}\right]^2} ,
\label{MCLM}
\end{equation}
where $dN/dz$ is the {\it observed} differential redshift number count
of the objects, $b_{\rm eff}(z)$ is the effective biasing factor for
the objects, and $\xi(r,z)$ is the {\it mass} two-point correlation
function at a comoving separation $r$. 

While the above expression looks physically reasonable, it is
important to properly {\it define} the two-point correlation function
on the light-cone hypersurface and then to derive useful and practical
expressions with clarifying the underlying approximations and
assumptions. Even apart from such a theoretical motivation, equation
(\ref{MCLM}) involves a double integration over the redshift which is
not easy to evaluate numerically. Thus it is desirable to find a more
practical approximate formula, if any.

The primary purpose of this paper is to present a rigorous theoretical
formulation to define and compute a two-point correlation function of
cosmological objects which explicitly takes into account the
light-cone effect for the first time (\S 2). We propose an expression
(eq.[\ref{eq:xiA}]) as a practically useful formula relevant to the
discussion of clustering of high-z objects at large separations (\S
3). As a specific example of high-redshift objects on the light-cone,
we first consider the Durham/AAT QSO samples (Shanks \& Boyle 1994;
Croom \& Shanks 1996) in \S 4.1.  We compute the corresponding
two-point correlation function on the basis of our formula
(eq.[\ref{eq:xiA}]), which turns out to be in good agreement with the
results of Matarrese et al. (1997) which adopts equation (\ref{MCLM}).
Unfortunately we are not able to make further quantitative comparison
with their predictions due to the fact that equation (\ref{MCLM}) is
hard to evaluate numerically. Therefore we illustrate the importance
of the light-cone effect by comparing with predictions based on
another {\it intuitive} expression (eq.[\ref{eq:defxiB}]) which we
propose as a counter example. For that purpose, we start from the QSO
luminosity function by Wallington \& Narayan (1993), and predict the
two-point correlation functions corresponding to the future SDSS and
2dF QSO samples (\S 4.2). Also we briefly discuss the dependence on
the underlying cosmological parameters (\S 4.3).

We should note here that the present paper is rather theoretical at
this stage. While we propose equation (\ref{eq:xiA}) as a well-defined
and practically useful expression for the light-cone correlation
function, the crucial difference shows up only in the regime where the
correlation is very weak which is hardly probed accurately with the
current data like Durham/AAT sample, for instance. Furthermore we have
not yet successfully included several important effects
observationally; redshift-distortion and evolution of bias among
others. We plan to report the further progress in those issues
elsewhere in due course (Nishioka \& Yamamoto 1999; Yamamoto \& Suto
1999). Such limitations are discussed in \S 5 along with our main
conclusions of the present paper. Throughout this paper we use the
units in which the light velocity $c$ is unity.

\section{Two-point clustering statistics on a light-cone hypersurface}

In what follows, we focus on the spatially-flat Friedmann -- Robertson
-- Walker space-time for simplicity, whose line element is expressed
in terms of the conformal time $\eta$ as
\begin{equation}
  ds^2 = a^2(\eta) \left[-d\eta^2+dr^2+r^2 d\Omega_{(2)}^2 \right] .
\label{metric}
\end{equation}
Since our fiducial observer is located at the origin of the
coordinates ($\eta=\eta_0$, $r=0$), an object at $r$ and $\eta$ on the
light-cone hypersurface of the observer satisfies a simple relation of
$r=\eta_0-\eta$.  While this is mainly why we adopt the spatially-flat
model, current observations indeed seem to favor such a cosmological
model (e.g., Garnavich et~al. 1998).

Denote the comoving number density of observed objects at $\eta$ and
$\bfx=(r,\bft)$ by $n(\eta,\bfx)$, then the corresponding number
density defined on the light-cone is written as
\begin{equation}
  \nLC(r,\bft)=n(\eta_0-r,r,\bft) .
\label{eq:nlc1}
\end{equation}
If we introduce the mean {\it observed} number density (comoving) and
the density fluctuation at $\eta$, $n_0(\eta)$ and
$\Delta(\eta,\bfx)$, on the constant-time hypersurface:
\begin{equation}
  n(\eta,\bfx) = n_0(\eta) \left[1+\Delta(\eta,\bfx)\right],
\end{equation}
equation (\ref{eq:nlc1}) is rewritten as
\begin{equation}
  \nLC(r,\bft)=n_0(\eta_0-r) \left[1+\Delta(\eta_0-r,r,\bft) \right] .
\label{nLC}
\end{equation}
Note that the mean {\it observed} number density $n_0(\eta)$ is
different from the mean number density of the objects
$\overline{n}(\eta)$ at $\eta$ by a factor of the selection function
$\phi(\eta)$ which depends on the luminosity function of the objects
and thus the magnitude-limit of the survey, for instance:
\begin{equation}
  n_0(\eta) = \overline{n}(\eta) \phi(\eta).
\label{eq:n0}
\end{equation}

When the observed density field of objects on the light-cone,
$\nLC(r,\bft)$ is given, one may compute the following two-point
statistics:
\begin{equation}
  {\cal X}(R)={1\over V^{\rm LC}}
  \int{d\Omega_{\hat \bfR}\over 4\pi} 
  \int dr_1 r_1^2 d\Omega_{\bft_1} 
  \int dr_2 r_2^2 d\Omega_{\bft_2}
  \nLC(r_1,\bft_1)\nLC(r_2,\bft_2)
  \delta^{(3)}(\bfx_1-\bfx_2-\bfR),
\label{C3}
\end{equation}
where $\bfx_1=(r_1,r_1\bft_1)$ and $\bfx_2=(r_2,r_2\bft_2)$ and
$R=|\bfR|$, $\hat \bfR=\bfR/R$, and $V^{\rm LC}$ is the comoving
survey volume of the data catalogue:
\begin{equation}
  V^{\rm LC} =\int_\rmin^\rmax  r^2 dr 
    \int d\Omega_\bft={4\pi \over 3} (\rmax^3 - \rmin^3) ,
\end{equation}
with $\rmax = r(z_{\rm max})$ and $\rmin= r(z_{\rm min})$ being the
boundaries of the survey volume. Although the second equality as well
as the analysis below assumes that the survey volume extends $4\pi$
steradian, all the results below can be easily generalized to the case
of the finite angular extension.

Substituting equation (\ref{nLC}), the ensemble average of an
estimator ${\cal X}(R)$ is explicitly written as
\begin{equation}
\bigl<{\cal X}(R)\bigr>={\cal U}(R)+{\cal W}(R),
\label{eq:xuw}
\end{equation}
where
\begin{eqnarray}
  &&{\cal U}(R)=
  {1\over V^{\rm LC}}
  \int{d\Omega_{\hat \bfR}\over 4\pi} 
  \int dr_1 r_1^2 \int d\Omega_{\bft_1} 
  \int dr_2 r_2^2 \int d\Omega_{\bft_2}
\nonumber
\\
  &&\hspace{1.5cm}
  \times
  n_0(\eta_0-r_1)n_0(\eta_0-r_2)
  \delta^{(3)}(\bfx_1-\bfx_2-\bfR),
\label{C33}
\end{eqnarray}
and
\begin{eqnarray}
  &&{\cal W}(R)=
  {1\over V^{\rm LC}}
  \int{d\Omega_{\hat \bfR}\over 4\pi} 
  \int dr_1 r_1^2 \int d\Omega_{\bft_1} 
  \int dr_2 r_2^2 \int d\Omega_{\bft_2}
  n_0(\eta_0-r_1)n_0(\eta_0-r_2)
\nonumber
\\
  &&\hspace{1.5cm}
  \times
  \Bigl<\Delta(\eta_0-r_1,r_1,\bft_1)\Delta(\eta_0-r_2,r_2,\bft_2)\Bigr>
  \delta^{(3)}(\bfx_1-\bfx_2-\bfR).
\label{C4}
\end{eqnarray}

Consider first ${\cal W}(R)$ which contains all the information of the
clustering. We show in Appendix A that in linear theory ${\cal W}(R)$
reduces to
\begin{eqnarray}
  {\cal W}(R)&=&
  {1\over V^{\rm LC }} {1\over \pi R}
  \int\int_{\cal S} dr_1 dr_2  r_1 r_2 \prod_{j=1}^2 
  \left[ n_0(\eta_0-r_j) D_1(\eta_0-r_j)\right]
\nonumber
\\
  &&\times \int dk k^2  P(k)  j_0(kR) b(k;\eta_0-r_1) b(k;\eta_0-r_2),
  \label{B67}
\end{eqnarray}
where $D_1(\eta)$ is the linear growth rate normalized to unity at
present ($\eta_0$), $P(k)$ is the power spectrum of the mass
fluctuations at $\eta_0$, $b(k;\eta)$ is the $k$-dependent linear bias
factor, $j_0(x)$ is the spherical Bessel function of the 0-th order, and
$\cal S$ denotes the region $|r_1-r_2|\leq R\leq r_1+r_2$.

Since we are generally interested in the case of $R\ll\rmax$, we can
use the approximation:
\begin{eqnarray}
  \int\int_{\cal S} dr_1 dr_2\simeq 
  \int_\rmin^\rmax dr_1 \int_{|R-r_1|}^{R+r_1} dr_2 .
\end{eqnarray}
Furthermore using that $\eta_0-r_2 \simeq \eta_0-r_1$, equation
(\ref{B67}) reduces to
\begin{eqnarray}
  &&{\cal W}(R)\simeq
  {4\pi\over V^{\rm LC }} 
  \int_\rmin^\rmax r^2 dr \left[ n_0(\eta_0-r) D_1(\eta_0-r) \right]^2
\nonumber
\\
  &&\hspace{1.5cm}\times
  {1\over 2\pi^2}\int  k^2 dk  P(k)  j_0(kR) 
                      \left[b(k;\eta_0-r)\right]^2 .
  \label{C69}
\end{eqnarray}
Therefore in linear theory, we can write
\begin{eqnarray}
  &&{\cal W}(R)\simeq{4\pi\over V^{\rm LC }} 
  \int_\rmin^\rmax r^2 dr \left[n_0(\eta_0-r)\right]^2 
                    \xi(R;\eta_0-r)_{\rm Source}~,
  \label{B69}
\end{eqnarray}
where $\xi(R;\eta)_{\rm Source}$ is the conventional two-point
correlation defined on the constant hypersurface at the source's
position:
\begin{eqnarray}
  &&\xi(R;\eta)_{\rm Source}=
  {1\over 2\pi^2} \int  k^2 dk P(k)  j_0(kR) 
                 \left[b(k;\eta) D_1(\eta)\right]^2~.
\label{eq:xislinear}
\end{eqnarray}

Consider next ${\cal U}(R)$. Repeating the similar calculation from
equation (\ref{C4}) to (\ref{B67}), equation (\ref{C33}) reduces to
\begin{eqnarray}
  {\cal U}(R)&=&
  {1\over V^{\rm LC }} {2\pi\over R}
  \int\int_{\cal S} dr_1 dr_2  r_1 r_2 
   n_0(\eta_0-r_1)n_0(\eta_0-r_2) ,
  \label{C80}
\end{eqnarray}
and the same approximation from equation (\ref{B67}) to (\ref{B69})
yields
\begin{eqnarray}
  {\cal U}(R)&\simeq&
  {4\pi\over V^{\rm LC }} \int_\rmin^{\rmax}   r^2 dr
   \left[n_0(\eta_0-r)\right]^2 .
  \label{C81}
\end{eqnarray}
Thus ${\cal U}(R)$ is independent of $R$ for $R\ll\rmax$, as expected.
        
Although the rigorous derivation in Appendix A is carried out entirely
in the framework of linear theory, the final result (\ref{B69}) would
be valid also for non-linear regimes; replacing equation
(\ref{eq:xislinear}) by its nonlinear counterpart, one may approximate
the correlation in equation (\ref{C4}) as
\begin{equation}
  \Bigl<\Delta(\eta_0-r_1,r_1,\bft_1)\Delta(\eta_0-r_2,r_2,\bft_2)\Bigr>
  \simeq\xi(r_1,r_2)\simeq\xi(R;\eta_0-r_{\rm h}),
\label{approxnl}
\end{equation}
where $r_{\rm h}=(r_1+r_2)/2$, ~$R=|r_1-r_2|$, and $\xi(R;\eta)$ is
the two-point correlation function on an equal-time hypersurface.  In
equation (\ref{approxnl}), we have neglected the angular dependence in
$\bigl<\Delta(\eta_0-r_1,r_1,\bft_1)\Delta(\eta_0-r_2,r_2,\bft_2)\bigr>$
and regarded it as a function of $r_1$ and $r_2$.  If this is the
case, the similar calculation to derive equation (\ref{C80}) yields
\begin{eqnarray}
  {\cal W}(R)&=&
  {1\over V^{\rm LC }} {2\pi\over R}
  \int\int_{\cal S} dr_1 dr_2  r_1 r_2 
   n_0(\eta_0-r_1)n_0(\eta_0-r_2)\xi(R;\eta_0-r_{\rm h}).
  \label{C90}
\end{eqnarray}
Again, with the same approximation from equation (\ref{B67}) to
(\ref{B69}), we have
\begin{eqnarray}
  {\cal W}(R)&\simeq&
  {4\pi\over V^{\rm LC }} \int_\rmin^{\rmax}   r^2  dr
   \left[n_0(\eta_0-r)\right]^2 \xi(R;\eta_0-r) ,
  \label{C91}
\end{eqnarray}
which is identical to equation (\ref{B69}) except that $\xi(R;
\eta_0-r)$ is not restricted to linear theory prediction.
Unfortunately we have not been able to justify the validity of the
approximation (\ref{approxnl}) strictly, but we expect that the above
qualitative and intuitive argument supports it. Note that this
approximation is also implicitly assumed in equation (\ref{MCLM}).
More importantly the resulting non-linear effect, which is taken into
account in this way in equation (\ref{C91}), is small as long as the
QSO clustering at $R\simgt 1h^{-1}{\rm Mpc}$ is considered as we will
explicitly show in \S 4.

\section{Definitions of the two-point correlation function on the light-cone}

In the previous section, we derived an approximate expression for the
two-point statistics $\left<{\cal X}(R)\right>$ and ${\cal W}(R)$ on
the light-cone. The remaining task is to define the corresponding
two-point correlation function on the light-cone.  A straightforward
application of equation (\ref{eq:xuw}) implies the definition:
\begin{equation}
  \xi_A^{\rm LC}(R) \equiv {\bigl<{\cal X}(R)\bigr>-{\cal U}(R)
  \over {\cal U}(R)}={{\cal W}(R)\over {\cal U}(R)}~. 
\label{eq:defxiA} 
\end{equation}
Substituting equations (\ref{B69}) and (\ref{C81}), the above
definition reduces to
\begin{equation}
  \xi_A^{\rm LC}(R) ={\displaystyle {\int_\rmin^\rmax dr r^2 
  n_0(\eta_0-r)^2 \xi(R;\eta_0-r)_{\rm Source}}
  \over
  {\displaystyle \int_\rmin^\rmax dr r^2 
  n_0(\eta_0-r)^2}}~.
\label{eq:xiA}
\end{equation}
Another possibility, which might look more similar to a conventional
pair-count estimator adopted in analyzing an observational map on a
constant-time hypersurface ($z\sim0$), is
\begin{equation}
  \xi_B^{\rm LC}(R)
 ={\bigl<{\cal X}(R)\bigr>-\bigl<{\barnLC}\bigr>^2
  \over \bigl<{\barnLC}\bigr>^2}=
  {{\cal U}(R)+{\cal W}(R)-\bigl<\barnLC\bigr>^2
  \over \bigl<\barnLC\bigr>^2} ,
\label{eq:defxiB}
\end{equation}
where $\barnLC$ is the mean number density on the light-cone
hypersurface and $\bigl<\barnLC\bigr>$ denotes its ensemble average:
\begin{eqnarray}
  \barnLC &=&{1\over \VLC}\int  r^2 dr \int d\Omega_{\bft}
  ~\nLC(r,\bft) ,\\
  \bigl<\barnLC\bigr> &=& {4\pi\over \VLC} \int_\rmin^\rmax  r^2 dr ~
  n_0(\eta_0-r) .
\end{eqnarray}

Two-point correlation functions on the light-cone can be computed from
the given catalogue of objects according to either {\it theoretical}
definition in a fairly straightforward manner, although one has to
assume a set of cosmological parameters {\it a priori} to translate
the observable coordinates to the comoving ones; first average over
the angular distribution and estimate the differential redshift number
count $dN/dz$ of the objects. Second distribute random particles over
the whole sample volume so that they obey the same $dN/dz$. Then the
conventional pair-count between the objects and random particles
yields ${\cal X}(R)$ (although not $\bigl<{\cal X}(R)\bigr>$, of
course), while ${\cal U}(R)$ is estimated from the pair-count of the
random particles themselves.  Finally $\barnLC$ can be computed as
$\int_\zmin^\zmax dz (dN/dz)$ with an appropriate normalization
constant.  In addition, when the observable comoving number density
$n_0(\eta)$ is independent of time, $\xi_A^{\rm LC}(R)$ and
$\xi_B^{\rm LC}(R)$ become identical and reduce to
\begin{equation}
 \xi_A^{\rm LC}(R)=\xi_B^{\rm LC}(R)
 ={\displaystyle {\int_\rmin^\rmax dr r^2 
  \xi(R;\eta_0-r)_{\rm Source}}
  \over
  {\displaystyle \int_\rmin^\rmax dr r^2 }}~.
\end{equation}
If the correlation function of objects does not evolve, i.e.,
$\xi(R;\eta_0-r)_{\rm Source} = \xi(R;\eta_0)_{\rm Source}$, however,
equation (\ref{eq:defxiA}) readily yields
\begin{equation}
\xi_A^{\rm LC}(R) = \xi(R;\eta_0)_{\rm Source}
\label{eq:xiAS}
\end{equation}
while equation (\ref{eq:defxiB}) does not.  In fact, as is shown in
\S~4.2, $\xi_B^{\rm LC}(R)$ does not vanish even where the real
spatial correlation is negligible. This is an example of a
non-physical effect due to the contamination by the intrinsic
evolution of mean number density of objects on the light-cone
hypersurface.

From a theoretical point of view, one might expect that a more
accurate expression on the light-cone should involve the double
integrals with respect to $z$ (e.g., eq. [\ref{MCLM}]) since the
spatial correlation along the line-of-sight should manifest itself in
the correlation between different redshifts. While this argument is
correct in a strict sense, we have shown that our expressions without
the double integral already take into account the leading term
properly. Thus unless one needs a next-order correction, our
approximate formulae are practically sufficient, and better suited for
numerical evaluations, compared with equation (\ref{MCLM}).

In summary, we propose equation (\ref{eq:defxiA}), rather than
equation (\ref{eq:defxiB}), as the best estimator of the two-point
correlation function on the light-cone hypersurface which captures the
true physics of correlation due to clustering. Nevertheless, it is
instructive to compare with the latter and discuss their difference in
some detail in order to understand some aspects of the light-cone
effect, as we show below.

\section{Application: two-point correlation functions 
  of high-redshift quasar samples on the light-cone}

\subsection{Durham/AAT QSO sample}

First we examine in details the extent to which equations
(\ref{MCLM}), (\ref{eq:xiA}), and (\ref{eq:defxiB}) lead to
quantitatively different predictions.  For this purpose, we compute
the correlation functions of Durham/AAT QSO sample (Shanks \& Boyle
1994; Croom \& Shanks 1996) in the Einstein -- de Sitter universe,
following Matarrese et al.(1997).  In this case,
$a(\eta)=(\eta/\eta_0)^2$, and the present conformal time $\eta_0$ is
$2/H_0$ where $H_0=100h$ km/s/Mpc is the Hubble constant.

According to Matarrese et~al. (1997), we adopt a simple assumption
that the quasar correlation function is given by the mass two-point
correlation function at that epoch multiplied by the bias factor which
depends on the mass of the hosting dark matter halo (e.g., Mo \& White
1996; Mo, Jing, \& White 1997; Jing 1998; see also Fang \& Jing 1998).
In this case, Matarrese et~al. (1997) found that the effective bias
$b_{\rm eff}(z)$ integrated over the halo mass larger than $M_{\rm
  min}$ is well-fitted to
\begin{equation}
  b_{\rm eff}(z) = 0.41 + [b_{\rm eff}(z=0)-0.41](1+z)^\beta,
\label{eq:beff}
\end{equation}
and obtained the values of $b_{\rm eff}(z=0)$ and $\beta$ for given
$M_{\rm min}$ in the standard cold dark matter (CDM) model (see their
Table 1). As for the mass two-point correlation function, we use the
nonlinear fitting formula by Peacock \& Dodds (1994,1996) for the CDM
power spectrum normalized to the cluster number counts (Kitayama \& Suto
1997; Kitayama, Sasaki \& Suto 1998).

\begin{figure}
\begin{center}
    \leavevmode\psfig{figure=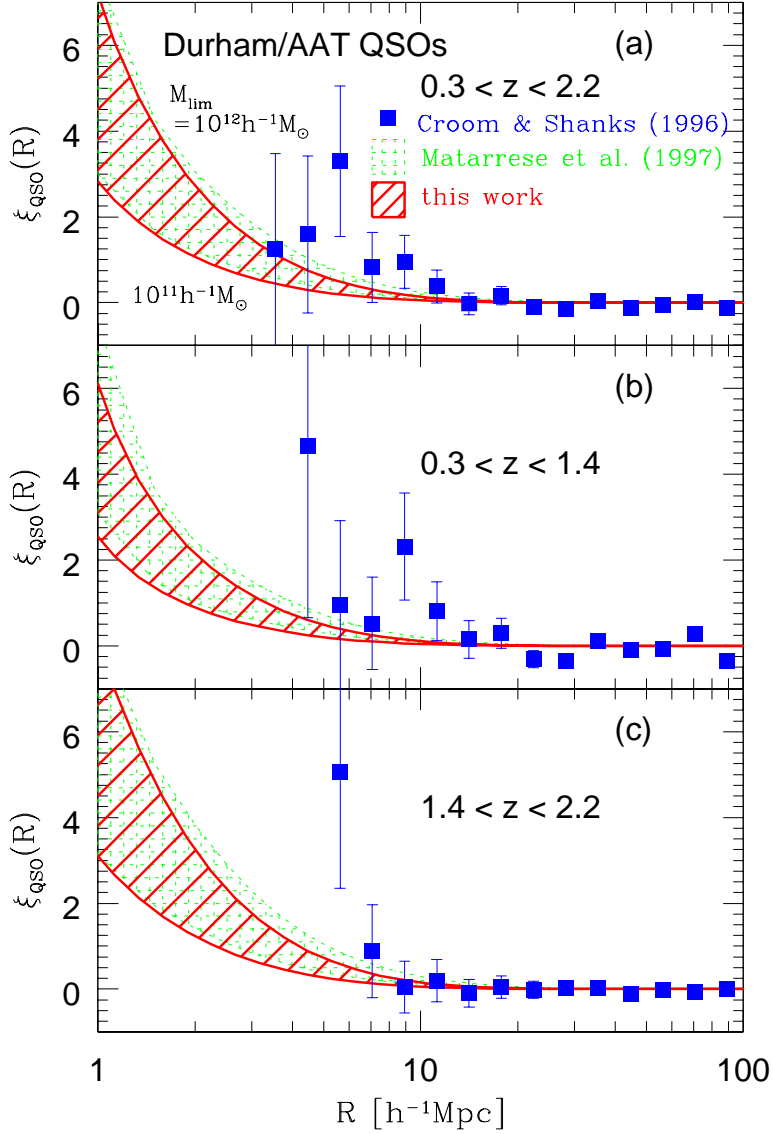,width=10cm}
\end{center}
\caption{ Two-point correlation functions of QSOs for Durham/AAT sample
  (Shanks \& Boyle 1994; Croom \& Shanks 1996) for (a) $0.3<z<2.2$,
  (b) $0.3<z<1.4$, and (c) $1.4<z<2.2$.  Our $\xi_A^{\rm LC}(R)$ with
  $M_{\rm lim}=10^{11}$ and $10^{12} h^{-1}M_\odot$ are plotted in the
  shaded region. (Our $\xi_B^{\rm LC}(R)$ is almost indistinguishable
  from $\xi_A^{\rm LC}(R)$ in this case.)  For comparison, the results
  by Matarrese et~al. (1997) are plotted in the dotted hatched
  regions.
\label{fig:xidurham}}
\end{figure}

We adopt the polynomial fit of Shanks \& Boyle (1994) to the
differential number count of the Durham/AAT QSO sample (the
corresponding B-band limiting magnitudes in several fields range 20.12
to 21.27):
\begin{equation}
\left({dN \over dz}\right)_{\rm Durham} 
= 2.738 - 57.573 z + 341.720 z^2 -548.737 z^3 
  + 408.908 z^4 - 145.698 z^5 + 19.989 z^6
\label{eq:dndzdurham}
\end{equation}
for $0.3<z<2.2$ (392 QSOs in total) which is plotted in Figure
\ref{fig:nzqso}a. We translate the count $N(z)$ to the observed mean
number density $n_0(r)$ according to
\begin{equation}
n_0(r) \propto {dN \over dz} {1\over r^2(z)} {dz \over dr} ,
\end{equation}
where the proportional constant is not necessary in our formulae.  In
this case we found that both equations (\ref{eq:xiA}) and
(\ref{eq:defxiB}) yield almost indistinguishable results.  Our results
are plotted in Figure \ref{fig:xidurham} along with those of Matarrese
et al. (1997). It is clear that the three expressions for the
light-cone correlation functions result in negligible difference, at
least comparing with both the statistical errors from the Durham/AAT
QSO sample and the theoretical uncertainty of the evolution of bias.
In the next subsection, however, we show that equations (\ref{eq:xiA})
and (\ref{eq:defxiB}) exhibit significantly behavior at large $R$
where the correlation is weak. Unfortunately we are not able to make
further comparison with equation (\ref{MCLM}) proposed by Matarrese et
al. (1997) since we are not sure how to deal with the double
integration precisely.

\subsection{2dF and SDSS QSO samples}

\begin{figure}
\begin{center}
     \leavevmode\psfig{figure=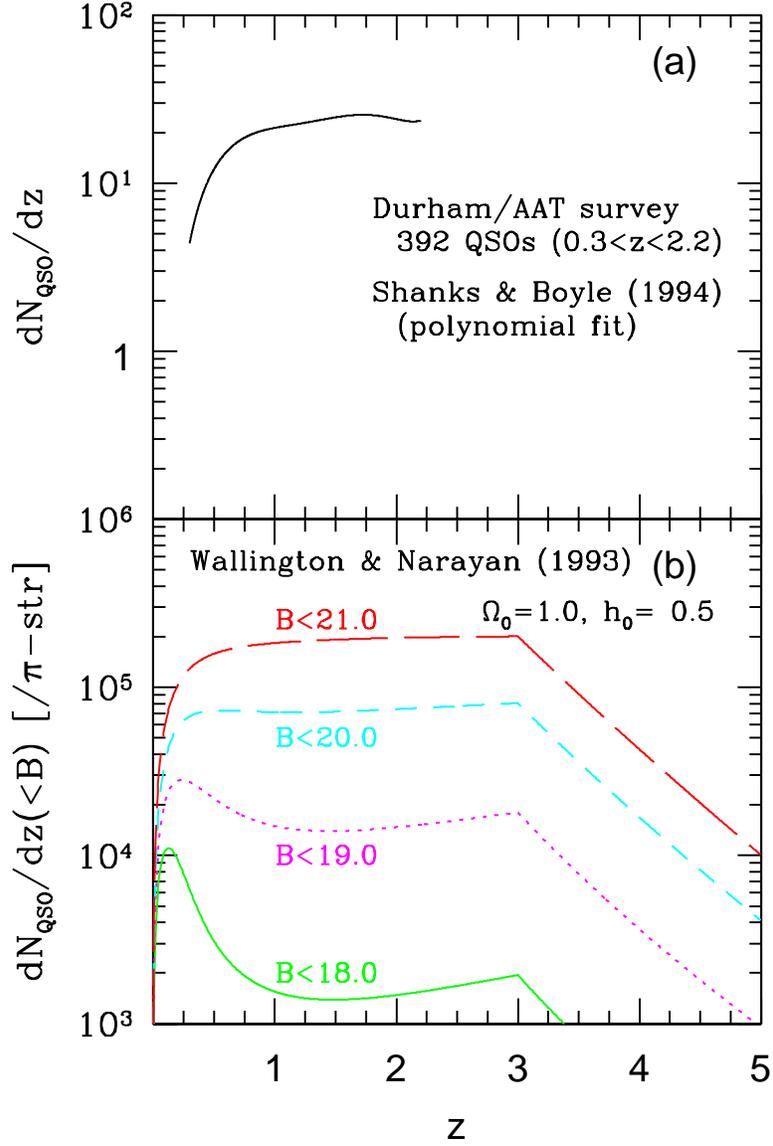,width=10cm}
\end{center}
\caption{Redshift distribution of QSOs from (a) the Durham/AAT sample
  (Shanks \& Boyle 1994) and that based on (b) the luminosity function
  by Wallington \& Narayan (1993) with different B-band limiting
  magnitudes.
\label{fig:nzqso}}
\end{figure}

It seems premature to draw
any cosmological conclusions from the comparison with the currently
available data samples.  Therefore we apply our formulae, $\xi_A^{\rm
  LC}(R)$ and $\xi_B^{\rm LC}(R)$, to predict the correlation
functions for the ongoing SDSS and 2dF QSO catalogues. In this case we
need a redshift dependent QSO luminosity function which is not
well-established, especially at higher redshifts. In what follows we
adopt the B-band quasar luminosity function according to Wallington \&
Narayan (1993; see also Boyle, Shanks \& Peterson 1988 and Nakamura \&
Suto 1997). To be specific, for $0.3<z<3$
\begin{eqnarray} 
\Phi(M_{\rm B},z) &=& {\Phi_* \over
10^{0.4(\alpha+1)[M_{\rm B}-M_{\rm B}^*(z)]} + 10^{0.4(\beta+1)[M_{\rm
B}-M_{\rm B}^*(z)]} } , \label{eq:phiqso} \\ 
M_{\rm B}^*(z) &=& M_{\rm B}^* - 2.5 k_{\rm L} \log(1+z) , 
\end{eqnarray}
with $M_{\rm B}^*= -20.91+5\log h$, $k_{\rm L}=3.15$, $\alpha=-3.79$,
$\beta=-1.44$, $\Phi_* = 6.4\times 10^{-6} h^3{\rm Mpc}^3$. For $z>3$,
they adopt
\begin{eqnarray} 
\Phi(M_{\rm B},z) &=& {\Phi_* \times
10^{-[A+0.4B(\beta+1)]} \over 10^{0.4(\alpha+1)[M_{\rm B}-M_{\rm
B}^*(z)]} + 10^{0.4(\beta+1)[M_{\rm B}-M_{\rm B}^*(z)]} } ,
\label{eq:phiqso3} 
\\ M_{\rm B}^*(z) &=& M_{\rm B}^* - 2.5 k_{\rm L} \log 4 + B, 
\end{eqnarray}
with $A=(z-3)\log 3.2$, $B=2.5A/(\alpha-\beta)$.  To compute the
B-band apparent magnitude from a quasar of absolute magnitude $M_{\rm
  B}$ at $z$ (with the luminosity distance $d_{\rm L}(z) =
(2/H_0)(1+z-\sqrt{1+z})$ in the Einstein -- de Sitter universe), we
applied the K-correction:
\begin{equation}
  B = M_{\rm B} + 5 \log(d_{\rm L}(z)/ 10 {\rm pc}) 
- 2.5(1-p)\log (1+z) 
\end{equation}
for the quasar energy spectrum $L_\nu \propto \nu^{-p}$ (we use
$p=0.5$). We adopt the B-band limiting magnitudes for 2dF and SDSS QSO
samples as $B_{\rm lim} = 20.85$ and 20, respectively.
The redshift distribution of QSOs predicted from the above luminosity
function is plotted in Figure \ref{fig:nzqso}b with the B-band limiting
magnitudes $B_{\rm lim}=(18\sim 21)$.

\begin{figure}
\begin{center}
   \leavevmode\psfig{figure=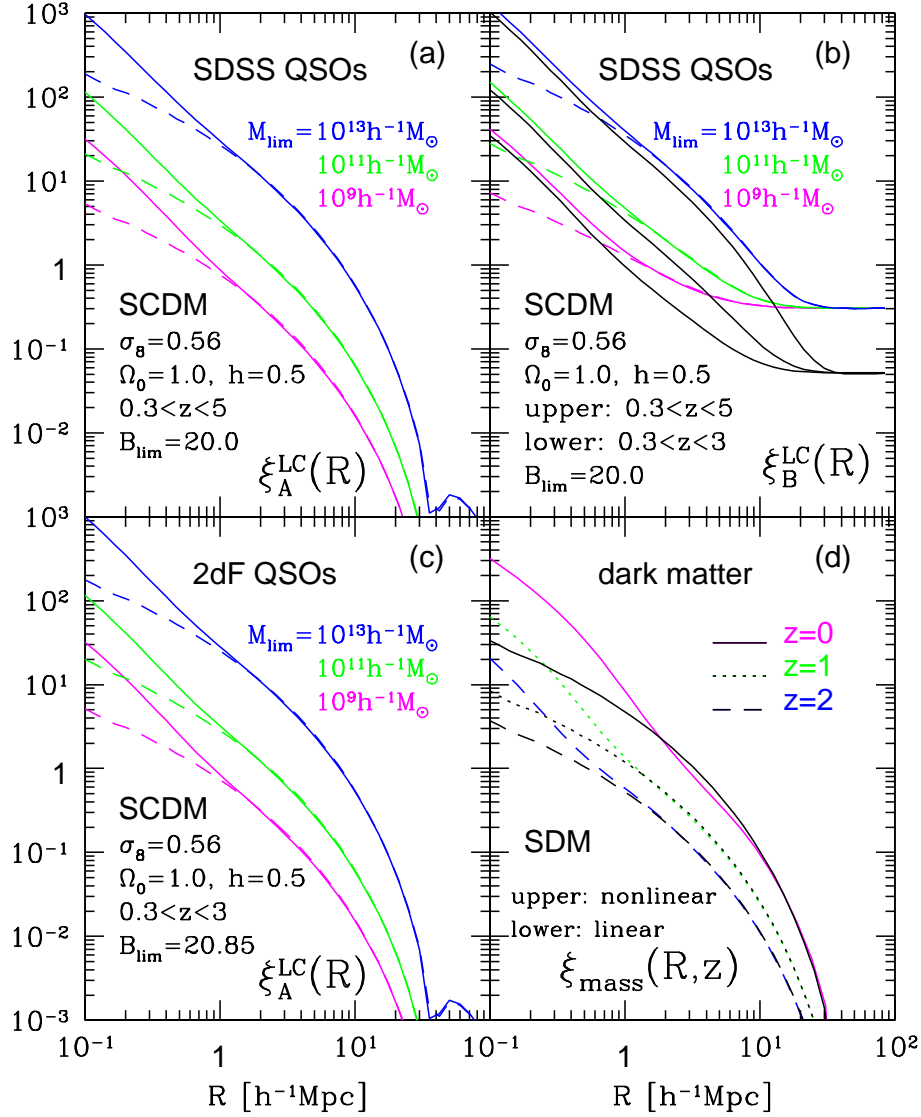,width=12cm}
\end{center}
\caption{ Two-point correlation functions of QSOs defined on the
  light-cone hypersurface in cluster normalized standard CDM models
  ($\sigma_8=0.56$). (a) $\xi_{A}(R)$ for the SDSS QSO catalogue for
  the threshold mass $M_{\rm lim}=10^9$, $10^{11}$ and $10^{13} h^{-1}
  M_\odot$. Nonlinear mass correlation function by Peacock \& Dodds
  (1996) is used for solid lines, while mass correlation function in
  linear theory is used for dashed lines; (b) the same as (a) for
  $\xi_{B}(R)$; (c) the same as (a) for the 2dF QSO catalogue; (d)
  linear (lower curves) and nonlinear (upper curves: Peacock \& Dodds
  1996) mass correlation functions defined on constant-time
  hypersurfaces $z=0$, 1 and 2.  \label{fig:xisdss}}
\end{figure}

We show our $\xi_A^{\rm LC}(R)$ and $\xi_B^{\rm LC}(R)$ for SDSS and
2dF QSO samples together with the nonlinear mass two-point correlation
functions in Figure \ref{fig:xisdss}.  Here we adopt the standard cold
dark matter (SCDM) model, in which $\Omega_0=1$, $\lambda=0.0$,
$h=0.5$, and the amplitude of the CDM density power spectrum is
normalized as $\sigma_8=0.56$ according to the cluster abundances
(e.g., Kitayama \& Suto 1996).  Three lines correspond to different
threshold mass $M_{\rm lim}=10^{13}$, $10^{11}$ and $10^9 M_\odot$
(from top to bottom) of the dark matter halos which are supposed to
host each QSO (Matarrese et~al. 1997).

We plot the results using the linear fluctuation power spectrum in
dashed lines while solid lines use the non-linear models of Peacock \&
Dodds (1996). The nonlinearity becomes important only on $R \simlt
1\himpc$ implying that one can safely ignore the nonlinear effect on
scales which are probed by (sparse) QSO samples in general.  The
linear and nonlinear correlation functions for dark matter (defined on
the constant-time hypersurface) are plotted in Figure \ref{fig:xisdss}d.

Panels (a) and (b) adopt the B-band limiting magnitude $B_{\rm
  lim}=20.0$ corresponding to the SDSS QSO catalogue, while panel (c)
adopts $B_{\rm lim}=20.85$ for the 2dF QSO catalogue. In practice,
those two predictions are very similar if we use $\xi_A^{\rm LC}(R)$.
On the other hand, $\xi_B^{\rm LC}(R)$ behaves quite differently; in
fact it levels off at large $R$, and the asymptotic value depends on
the range of $z$ (Fig.\ref{fig:xisdss}b). This is qualitatively
explained as follows; since ${\cal W}(R)$ decreases as $R$ and ${\cal
  U}(R)$ is independent of $R$ at large $R$, equations (\ref{C81}) and
(\ref{eq:defxiB}) imply that $\xi_B^{\rm LC}(R)$ becomes
asymptotically
\begin{equation}
 \lim_{R\rightarrow \infty}\xi_B^{\rm LC}(R)= 
{{\cal U}(R)-\bigl<\barnLC\bigr>^2  \over \bigl<\barnLC\bigr>^2}
= {{\displaystyle \int_\rmin^\rmax r^2dr \times 
          \int_\rmin^\rmax r^2dr [n_0(\eta_0-r)]^2 } 
\over
   {\displaystyle \left[\int_\rmin^\rmax r^2dr n_0(\eta_0-r) \right]^2 } 
} -1 .
\end{equation}
The right-hand side of the above equation does not vanish unless
$n_0(\eta)$ is constant, and becomes larger as $n_0(\eta)$ changes
more significantly within the survey volume. This is why $\xi_B^{\rm
  LC}(R)$ becomes significantly different from $\xi_A^{\rm LC}(R)$
when the survey limit exceeds $z \sim 3$ where the QSO number density
begins to decrease substantially (see Fig. \ref{fig:nzqso}a and b).
Incidentally this feature was not clear in Figure \ref{fig:xidurham}
because the Durham/AAT QSO sample is limited to $z<2.2$ and also
because the Figure is plotted in linear, not log, scale.

This clearly illustrates the fact that the spatial clustering of
objects defined on the light-cone could be apparently contaminated and
mixed up with the intrinsic evolution of mean density of objects and
even with the shape of observational selection function.  In this
respect, our proposal of $\xi^{\rm LC}_A(R)$ is a more reliable and
robust definition describing the two-point correlation on the
light-cone.

\subsection{dependence on the cosmological model parameters}

The next important question is the extent to which different
cosmological models lead to different predictions of two-point
correlations. For this study, we need the QSO luminosity function and
bias models for the arbitrary cosmological models which are not
available at this point. So we simply adopt the Durham/AAT QSO
redshift distribution function (\ref{eq:dndzdurham}) and the bias
model (\ref{eq:beff}), although the latter is derived in the Einstein --
de Sitter universe.

\begin{figure}
\begin{center}
    \leavevmode\psfig{figure=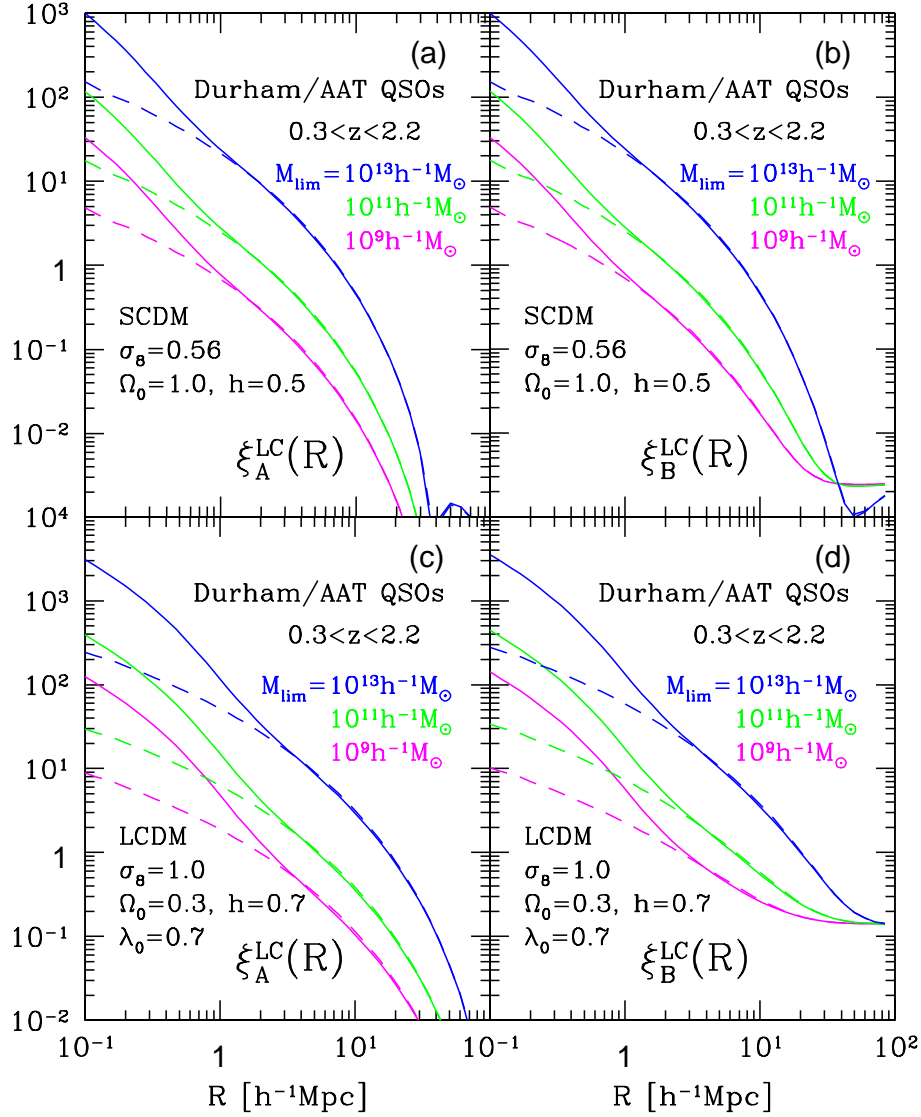,width=12cm}
\end{center}
\caption{ Dependence on the cosmological models. 
  (a) $\xi_{A}(R)$ for the SCDM model
  $(\Omega_0=1,~h=0.5,~\sigma_8=0.56)$ of QSO for Durham/AAT sample
  (Shanks \& Boyle 1994; Croom \& Shanks 1996).  Our $\xi_A^{\rm
    LC}(R)$ with $M_{\rm lim}=10^{9}$, $10^{11}$, and $10^{13} h^{-1}
  M_\odot$ for bias model are plotted. Nonlinear mass correlation
  function by Peacock \& Dodds (1996) is used for solid lines, while
  its linear theory counterpart is used for dashed lines; (b) the same
  as (a) for $\xi_{B}(R)$; (c) the same as (a) for LCDM model
  $(\Omega_0=0.3,~\lambda_0=0.7,~~h=0.7,~\sigma_8=1.0)$; (d) the same
  as (c) for $\xi_{B}(R)$.
\label{fig:xilambda}}
\end{figure}

\begin{figure}
\begin{center}
    \leavevmode\psfig{figure=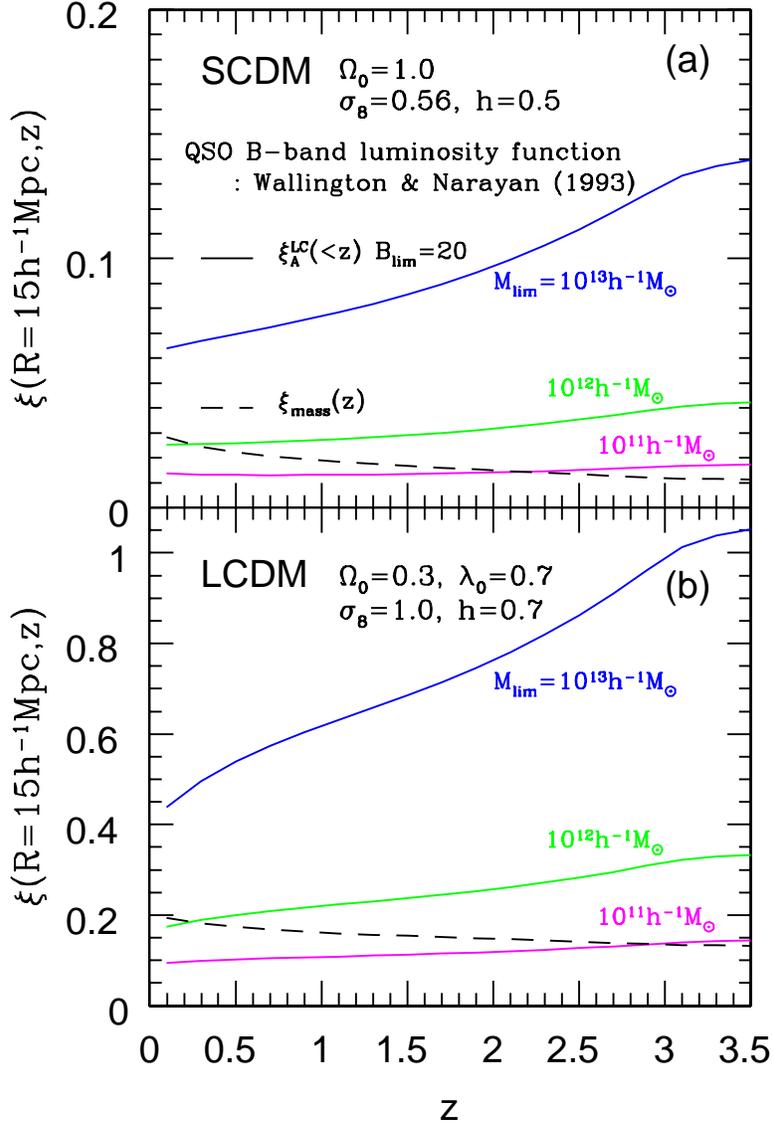,width=10cm}
\end{center}
\caption{ Evolution of amplitudes of two-point correlation functions at
 $R=15\himpc$ of QSOs on the light-cone in (a) SCDM, and
  (b) LCDM models.
\label{fig:xizqso}}
\end{figure}

In Figure \ref{fig:xilambda}, we compare the predictions in the SCDM
model against those in the spatially-flat low-density CDM (LCDM) model
in which $\Omega_0=0.3$, $\lambda=0.7$, $h=0.7$, and $\sigma_8=1.0$.
Figure \ref{fig:xizqso} plots the amplitude of $\xi^{\rm LC}_A(R)$ at
$R=15h^{-1}$Mpc as a function of the redshift of the survey limit
while keeping $B_{\rm lim}=20$. Since LCDM has more power on larger
scales and the fluctuation amplitude $\sigma_8$ is larger, the
correlation is stronger compared with that in SCDM given the same bias
model.  It should be noted that either model predicts that the QSO
correlation amplitude {\it increases} as $z$ becomes larger unlike the
mass correlation which always grows from high to low redshifts. This
qualitative feature is consistent with the finding of La Franca,
Andreani \& Cristiani (1998), and implies that the evolution of bias
dominates the growth of high-z objects in addition to the growth rate
of the mass fluctuations. Therefore this kind of comparison should
yield profound cosmological implications on the nature of QSOs,
although it would be premature to draw any decisive conclusion with
the current theoretical understanding of the bias (e.g., Fry 1996; Mo
\& White 1996; Mo. Jing, \& White 1997; Fang \& Jing 1998; Jing 1998)
and statistics of the observational samples.

\section{Discussion and conclusions}

Although the clustering of objects at high redshifts has been
extensively discussed in the literature, the previous theoretical
analysis was largely based on a more or less qualitative treatment of
the light-cone effect (Matarrese et al. 1997; Matsubara, Suto \&
Szapudi 1997; Nakamura, Matsubara \& Suto 1998). For on-going wide and
deep surveys like SDSS and 2dF QSO surveys, the light-cone effect
becomes very important either as a contamination of the real signal or
as a cosmological probe.

In the present paper, we developed a theoretical formulation which
properly takes account of the light-cone effect for the first time.
Strictly speaking we were able to derive our main result, equation
(\ref{B69}), only in linear theory, but we expect that the same
expression would be a good approximation even in nonlinear regime. In
any case the correlations of objects at high redshifts are described
almost entirely in linear theory for $R\simgt 1 \himpc$
(Figs.\ref{fig:xisdss} and \ref{fig:xilambda}), and in practice the
expression is guaranteed to be valid on the scales of interest.

We proposed a well-defined expression, equation (\ref{eq:xiA}), for
two-point correlation functions on light-cone, and compared its
implications with those of another possibility (\ref{eq:defxiB}).  As
long as the Durham/AAT QSO sample is considered, both expressions give
almost indistinguishable results and in fact they are also in good
agreement with the previous proposal by Matarrese et al. (1997).
Although we are not able to make further quantitative comparison with
the latter, this implies that our proposal without the double
integration over the redshift distribution is more practical in making
theoretical predictions.  Then we applied our expressions and computed
the two-point correlation functions on light-cone for future SDSS and
2dF QSO samples, and showed an example that the light-cone effect
could mix up the true spatial clustering of objects and the intrinsic
evolution of mean density of objects if one uses a native definition like
equation (\ref{eq:defxiB}).

In fact there are several issues which remain to be worked out in the
present context. The QSO luminosity function and the evolution of bias
play central roles in confronting observations and predictions of the
QSO correlation functions.  In this paper, we tentatively adopted the
expression by Wallington \& Narayan (1993) even in LCDM models,
although it is relevant only in the Einstein - de Sitter model.  It is
highly desirable that the QSO luminosity function in general
cosmological models is derived from future observations and becomes
available for the theoretical analysis. Theoretical approaches to
determine the time (and scale) dependence of bias are just in the
beginning (Fry 1996; Mo \& White 1996; Jing 1998; Dekel \& Lahav
1999). We did not attempt to explore a range of possible bias models,
but rather adopt a simple fit by Matarrese et al. (1997) as a specific
example. Most likely QSO correlation functions from future samples are
the most straightforward tool to test the several bias models in
further detail.  In the present paper, we have neglected the
redshift-space distortion either to the peculiar velocity of the
objects (Kaiser 1987; Hamilton 1997; Nishioka \& Yamamoto 1999) or to the
geometry of the universe (Matsubara \& Suto 1996; Ballinger, Peacock
\& Heavens 1996).  Definitely these effects should be crucial in the
quantitative comparison to high precision, and we plan to incorporate
the effect in future work. Nevertheless we hope that the current paper
presents a convincing case that the light-cone effect should be
properly taken into account in analysing the future surveys of
high-redshifts objects.

\bigskip
\bigskip

We deeply thank Sabino Matarrese and Lauro Moscardini for providing us
their results in a computer readable form and also for useful
correspondences on the formulation of two-point correlation functions
on the light-cone.  We are also grateful to Scott Croom and Tom Shanks
for allowing us to include their correlation data in Figure
\ref{fig:xidurham}, Y.P. Jing for providing the routines to compute
the nonlinear mass two-point correlation functions, and to Takahiro T.
Nakamura for discussions on the quasar luminosity function.  We thank
the referees, Richard Ellis and Stephen Landy, for constructive
comments on the earlier manuscript which helped improve the
presentation of the present paper.  K.Y. thanks Yasufumi Kojima for
comments. This research was supported in part by the Grants-in-Aid by
the Ministry of Education, Science, Sports and Culture of Japan
(09740203) and (07CE2002) to RESCEU.

\clearpage
\baselineskip=13pt
\parskip1pt
\centerline{\bf REFERENCES}
\bigskip

\def\apjpap#1;#2;#3;#4; {\pp#1, {#2}, {#3}, #4}
\def\apjbook#1;#2;#3;#4; {\pp#1, {#2} (#3: #4)}
\def\apjppt#1;#2; {\pp#1, #2.}
\def\apjproc#1;#2;#3;#4;#5;#6; {\pp#1, {#2} #3, (#4: #5), #6}

\apjpap Ballinger, W.E., Peacock, J.A., \& Heavens, A.F. 1996;MNRAS;282;877;
\apjppt Boyle, B.J., Croom, S.M., Smith, R.J., Shanks, T., Miller L., 
 \& Loaring, N. 1998;Phil.Trans.R.Soc.Lond.A, in press (astro-ph/9805140);
\apjpap Boyle, B.J., Shanks, T., \& Peterson, B.A. 1988;MNRAS;235;935;
\apjpap Carrera, F.J. et~al. 1998;MNRAS;299;229;
\apjpap Croom, S.M. \& Shanks, T. 1996;MNRAS;281;893;
\apjppt Dekel, A. \& Lahav, O. 1999;ApJ, in press (astro-ph/9806193);
\apjpap Fang, L.Z. \& Jing, Y.P. 1998;ApJ;502;L95;
\apjpap Fry, J. N. 1996;ApJ;461;L65;
\apjpap Garnavich, P.M. et~al. 1998;ApJ;493;L53;
\apjppt Hamilton, A.J.S. 1997; to appear in the Proceedings of
  Ringberg Workshop on Large-Scale Structure, edited by Hamilton, D.
  (astro-ph/9708102);
\apjpap Hamilton, A.J.S., Kumar, P., Lu, E., \& Matthews, A. 1991;
  ApJ;374;L1;
\apjpap Jain, B., Mo, H.J., \& White, S.D.M. 1995;MNRAS;276;L25;
\apjpap Jing, Y.P. 1998;ApJ;503;L9;
\apjpap Jing, Y.P.,  \& Suto, Y. 1998;ApJ;494;L5;
\apjpap Kaiser, N. 1987;MNRAS;227;1;
\apjpap Kitayama, T., Sasaki,S., \& Suto, Y. 1998;PASJ;50;1;
\apjpap Kitayama, T. \& Suto, Y. 1997;ApJ;490;557;
\apjpap La Franca, F., Andreani, P., \& Cristiani, S. 1998;ApJ;497;529;
\apjpap Lahav, O., Piran,T., \& Treyer, M. 1997;MNRAS;284;499;
\apjpap Magliocchetti. M.,  Maddox, S.J.,  Lahav, O.,  \& Wall, J.V 1998;
MNRAS;300;257;
\apjbook Magnus, W., Oberhettinger, F., \& Soni, R.P. 1966;Formulas and 
Theorems for the Special Functions of Mathematical 
Physics;Springer-Verlag;Berlin;
\apjpap Matarrese, S., Coles, P., Lucchin, F., \& Moscardini, L. 1997;
 MNRAS;286;115;
\apjpap Matsubara, T. \& Suto, Y. 1996;ApJ;470;L1;
\apjpap Matsubara, T. , Suto, Y., \& Szapudi 1997;ApJ;491;L1;
\apjpap Mo, H.J., Jing, Y.P., \& White, S.D.M. 1997;MNRAS;284;189;
\apjpap Mo, H.J., \& White, S.D.M. 1996;MNRAS;282;347;
\apjpap Nakamura, T.T., Matsubara, T. \& Suto, Y. 1998;ApJ;494;13;
\apjpap Nakamura, T.T.,  \& Suto, Y. 1997;Prog. Theor. Phys.;97;49;
\apjppt Nishioka, H, \& Yamamoto, K. 1999; ApJ submitted;
\apjpap Peacock, J.A. \& Dodds, S.J. 1994;MNRAS;267;1020;
\apjpap Peacock, J.A. \& Dodds, S.J. 1996;MNRAS;280;L19;
\apjpap Shanks, T. \& Boyle, B.J. 1994;MNRAS;271;753;
\apjpap Steidel, C.C., Adelberger, K.L., Dickinson, M., Giavalisco,
M., Pettini, M., \& Kellogg, M. 1998;ApJ;492;428;
\apjpap Wallington, S., \& Narayan, R. 1993;ApJ;403;517;
\apjppt Yamamoto, K., \& Sugiyama, N. 1998;Phys.Rev.D~58,~103508;
\apjppt Yamamoto, K., \& Suto, Y. 1999; in preparation;

\clearpage
\begin{appendix}

\section*{Appendices}

\bigskip

\section{Calculation of ${\cal W}(R)$}

We present an explicit derivation of equation (\ref{B67}) from
equation (\ref{C4}):
\begin{eqnarray}
  &&{\cal W}(R)=
  {1\over V^{\rm LC}}
  \int{d\Omega_{\hat \bfR}\over 4\pi} 
  \int dr_1 r_1^2 \int d\Omega_{\bft_1} 
  \int dr_2 r_2^2 \int d\Omega_{\bft_2}
  n_0(\eta_0-r_1)n_0(\eta_0-r_2)
\nonumber
\\
  &&\hspace{1.5cm}
  \times
  \Bigl<\Delta(\eta_0-r_1,r_1,\bft_1)\Delta(\eta_0-r_2,r_2,\bft_2)\Bigr>
  \delta^{(3)}(\bfx_1-\bfx_2-\bfR)
\label{app4}
\end{eqnarray}
in linear theory.

First let us expand the number density contrast $\Delta(\eta,r,\bft)$
as
\begin{equation}
    \Delta(\eta,r,\bft)=\int_0^\infty dk \sum_{l,m} \Delta_{klm}(\eta)
    {\cal Y}_{klm}(r,\bft),
\label{C2}
\end{equation}
in terms of the normalized harmonics:
\begin{equation}
   {\cal Y}_{klm}(r,\bft)= X_{kl}(r) Y_{lm}(\Omega_{\bft}),
\end{equation}
where
\begin{equation}
   X_{kl}(r)=\sqrt{{2\over \pi}} k j_l(kr),
\end{equation}
and $Y_{lm}(\Omega_{\bft})$ and $j_l(x)$ are the spherical harmonics
and the spherical Bessel function, respectively. 

Substituting equation (\ref{C2}) into equation (\ref{app4}), one
obtains
\begin{eqnarray}
  {\cal W}(R)&=&
  {1\over V^{\rm LC}}
  \int{d\Omega_{\hat \bfR}\over 4\pi} 
  \int dr_1 r_1^2 \int d\Omega_{\bft_1} 
  \int dr_2 r_2^2 \int d\Omega_{\bft_2}
  n_0(\eta_0-r_1)n_0(\eta_0-r_2)
\nonumber
\\
  &&\times
  \int dk_1 \sum_{l_1,m_1}   \int dk_2 \sum_{l_2,m_2} 
  \bigl<\Delta_{k_1l_1m_1}(\eta_0-r_1)
  \Delta_{k_2l_2m_2}^*(\eta_0-r_2) \bigr>
\nonumber
\\
  &&\times
  {X}_{k_1l_1}(r_1)Y_{l_1m_1}(\Omega_{\bft_1})
  {X}_{k_2l_2}(r_2)Y_{l_2m_2}^*(\Omega_{\bft_2})~
  \delta^{(3)}(\bfx_1-\bfx_2-\bfR).
\label{C6}
\end{eqnarray}

In Appendix B, we show that the ensemble average of the mode
coefficient in the integrand of equation (\ref{C6}) reduces to the
following function:
\begin{equation}
  \bigl<\Delta_{k_1l_1m_1}(\eta_0-r_1) 
  \Delta_{k_2l_2m_2}^*(\eta_0-r_2)\bigr>
  =Q(k_1,k_2,r_1,r_2,l_1) \delta_{l_1l_2}\delta_{m_1m_2},
\label{eq:q12}
\end{equation}
where $\delta_{l_1l_2}$ and $\delta_{m_1m_2}$ are the Kronecker's
delta. In addition, we use the relations 
\begin{equation}
  \delta^{(3)}(\bfx_1-\bfx_2-\bfR)={1\over (2\pi)^3}
  \int d^3\bfk ~e^{-i\bfk\cdot(\bfx_1-\bfx_2-\bfR)},
\end{equation}
and
\begin{equation}
  e^{-i\bfk\cdot\bfx}= 4\pi \sum_{l} \sum_{m=-l}^{l}
  (-i)^l j_{l}(k|\bfx|) Y_{lm}(\Omega_{\hat\bfk})
  Y_{lm}^*(\Omega_{\hat\bfx}),
\end{equation}
and then equation (\ref{C6}) becomes
\begin{eqnarray}
  {\cal W}(R)&=&
  {1\over V^{\rm LC}}
  \int{d\Omega_{\hat \bfR}\over 4\pi} 
  \int dr_1 r_1^2 \int d\Omega_{\bft_1} 
  \int dr_2 r_2^2 \int d\Omega_{\bft_2}
  n_0(\eta_0-r_1)n_0(\eta_0-r_2)
\nonumber
\\
  &&\times
  \int dk_1 \sum_{l_1,m_1}   \int dk_2 \sum_{l_2,m_2} 
  Q(k_1,k_2,r_1,r_2,l_1) \delta_{l_1l_2}\delta_{m_1m_2}
\nonumber
\\
  &&\times
  {X}_{k_1l_1}(r_1)Y_{l_1m_1}(\Omega_{\bft_1})
  {X}_{k_2l_2}(r_2)Y_{l_2m_2}^*(\Omega_{\bft_2})~
\nonumber
\\
  &&\times{1\over (2\pi)^3}\int d^3\bfk
  ~4\pi \sum_{L_1M_1} (-i)^{L_1} j_{L_1}(kr_1) Y_{L_1M_1}(\Omega_{\hat\bfk})
  Y_{L_1M_1}^*(\Omega_{\bft_1}) 
\nonumber
\\
  &&\hspace{2.3cm}
  \times
  4\pi \sum_{L_2M_2} (i)^{L_2} j_{L_2}(kr_2) Y_{L_2M_2}^*(\Omega_{\hat\bfk})
  Y_{L_2M_2}(\Omega_{\bft_2}) 
\nonumber
\\
  &&\hspace{2.3cm}
  \times
  4\pi \sum_{L_3M_3} (i)^{L_3} j_{L_3}(kR) Y_{L_3M_3}^*(\Omega_{\hat\bfk})
  Y_{L_3M_3}(\Omega_{\hat {\bf R}})~, 
\label{C61}
\end{eqnarray}
where $k=|\bfk|$ and $\hat \bfk=\bfk/|\bfk|$.
Integrating over $\Omega_{\bft_1}$, $\Omega_{\bft_2}$ and
$\Omega_{{\hat {\bf R}}}$ yields
\begin{eqnarray}
  {\cal W}(R)&=&
  {1\over V^{\rm LC}}
  \int dr_1 r_1^2 
  \int dr_2 r_2^2 
  n_0(\eta_0-r_1)n_0(\eta_0-r_2)
\nonumber
\\
  &&\times
  \int dk_1 \sum_{l_1,m_1}   \int dk_2 \sum_{l_2,m_2} 
  Q(k_1,k_2,r_1,r_2,l_1) \delta_{l_1l_2}\delta_{m_1m_2}
  {X}_{k_1l_1}(r_1)
  {X}_{k_2l_2}(r_2)
\nonumber
\\
  &&\times{(4\pi)^2 \over (2\pi)^3}\int d^3\bfk
  (-i)^{l_1-l_2} j_{l_1}(kr_1)j_{l_2}(kr_2)j_{0}(kR) 
  Y_{l_1m_1}(\Omega_{\hat\bfk}) Y_{l_2m_2}^*(\Omega_{\hat\bfk}).
\label{C62}
\end{eqnarray}
and the further integration over $\Omega_{\hat {\bfk}}$ gives
\begin{eqnarray}
  {\cal W}(R)&=&
  {1\over V^{\rm LC}}
  \int dr_1 r_1^2 
  \int dr_2 r_2^2 
  n_0(\eta_0-r_1)n_0(\eta_0-r_2)
\nonumber
\\
  &&\times
  \int dk_1 \int dk_2 \sum_{l} (2l+1)   
  Q(k_1,k_2,r_1,r_2,l) {X}_{k_1l}(r_1) {X}_{k_2l}(r_2)
\nonumber
\\
  &&\times{(4\pi)^2 \over (2\pi)^3}\int dk k^2
  j_{l}(kr_1)j_{l}(kr_2)j_{0}(kR). 
\label{C63}
\end{eqnarray}
Noting the relation (e.g., Magnus et~al. 1966):
\begin{eqnarray}
  \int dk k^2 j_{l}(kr_1)j_{l}(kr_2)j_{0}(kR)
  = \left\{
      \begin{array}{ll}
        {\pi\over 4r_1r_2 R}
  P_l\biggl({r_1^2+r_2^2-R^2\over 2r_1r_2}\biggr) &
        \mbox{($|r_1-r_2|< R< r_1+r_2$)}, \\ 
        0 &
        \mbox{($R< |r_1-r_2|$, $R>r_1+r_2 $)}, 
      \end{array}
   \right. 
\end{eqnarray}
we find
\begin{eqnarray}
  &&{\cal W}(R)=
  {1\over V^{\rm LC }} {1\over \pi R}
  \int\int_{\cal S} dr_1 dr_2  r_1 r_2 n_0(\eta_0-r_1)n_0(\eta_0-r_2)
  \int dk_1 \int dk_2 k_1 k_2 
\nonumber
\\
  &&\hspace{1cm}\times \sum_{l} (2l+1)   
  Q(k_1,k_2,r_1,r_2,l) j_l(k_1r_1) j_l(k_2r_2)
  P_l\biggl({r_1^2+r_2^2-R^2\over 2r_1r_2}\biggr),
\label{C65}
\end{eqnarray}
where $\cal S$ denotes the region $|r_1-r_2|\leq R\leq r_1+r_2$.

So far we did not assume anything on the density field, i.e.,
$Q(k_1,k_2,r_1,r_2,l)$ can be an arbitrary function.  If we consider
the case of linear theory and a scale-dependent but still local
biasing scheme, each mode is decoupled and we can write
\begin{eqnarray}
  Q(k_1,k_2,r_1,r_2,l)= 
  D_1(\eta_0-r_1) D_1(\eta_0-r_2)b(k_1;\eta_0-r_1)b(k_2;\eta_0-r_2)
  P(k)\delta(k_1-k_2),
\end{eqnarray}
where $D_1(\eta)$ is the linear growth rate normalized to unity 
at present and $b(k;\eta)$ is the scale-dependent bias factor.
In this case, equation (\ref{C65}) is written as
\begin{eqnarray}
  &&{\cal W}(R)=
  {1\over V^{\rm LC }} {1\over \pi R}
  \int\int_{\cal S} dr_1 dr_2  \prod_{j=1}^2
  \biggl(r_j n_0(\eta_0-r_j) D_1(\eta_0-r_j)\biggr)
\nonumber
\\
  &&\hspace{1cm}
  \times \int dk k^2  P(k) b(k;\eta_0-r_1) b(k;\eta_0-r_2)
\nonumber
\\
  &&\hspace{1cm}
  \times \sum_{l} (2l+1) 
  j_l(kr_1) j_l(kr_2)P_l\biggl({r_1^2+r_2^2-R^2\over 2r_1r_2}\biggr).
\label{C66}
\end{eqnarray}
Using the formula:
\begin{equation}
  \sum_{l} (2l+1) 
  j_l(kr_1) j_l(kr_2)P_l\biggl({r_1^2+r_2^2-R^2\over 2r_1r_2}\biggr)
  =j_0(kR),
\end{equation}
we finally obtain
\begin{eqnarray}
  {\cal W}(R)&=&
  {1\over V^{\rm LC }} {1\over \pi R}
  \int\int_{\cal S} dr_1 dr_2  r_1 r_2 \prod_{j=1}^2 
  \biggl( n_0(\eta_0-r_j) D_1(\eta_0-r_j)\biggr)
\nonumber
\\
  &&\times \int dk k^2  P(k)  j_0(kR) b(k;\eta_0-r_1) b(k;\eta_0-r_2),
  \label{C67}
\end{eqnarray}
which is identical to equation (\ref{B67}).

\bigskip

\section{The power spectrum of mode coefficients}

Here we derive equation (\ref{eq:q12}) which is used in Appendix
A. Since equation (\ref{C2}) implies that the mode coefficient
$\Delta_{klm}(\eta)$ is expressed as
\begin{equation}
    \Delta_{klm}(\eta)=\int dr r^2 d\Omega_{\bft}\Delta(\eta,r,\bft)
    {\cal Y}_{klm}(r,\bft) ,
\label{appB1}
\end{equation}
one obtains 
\begin{eqnarray}
  &&\bigl<\Delta_{k_1l_1m_1}(\eta_1) 
  \Delta_{k_2l_2m_2}^*(\eta_2)\bigr>=
  \int dr_1 r_1^2 d\Omega_{\bft_1}\int dr_2 r_2^2 d\Omega_{\bft_2}
\nonumber
\\
  &&\hspace{2cm}\times
  \bigl<\Delta(\eta_1,r_1,\bft_1)\Delta(\eta_2,r_2,\bft_2)\bigr>
    {\cal Y}_{k_1l_1m_1}(r_1,\bft_1) {\cal Y}^*_{k_2l_2m_2}(r_2,\bft_2) .
\label{appB2}
\end{eqnarray}
Since the correlation on the light-cone hypersurface does not have any
special direction in $\bft$, the ensemble average
$\bigl<\Delta(\eta_1,r_1,\bft_1)\Delta(\eta_2,r_2,\bft_2)\bigr>$ 
should be a function of $\eta_1$, $\eta_2$, $r_1$, $r_2$ and 
$\cos\theta (\equiv\bft_1\cdot\bft_2)$:
\begin{equation}
  \bigl<\Delta(\eta_1,r_1,\bft_1)\Delta(\eta_2,r_2,\bft_2)\bigr>
  =F(\eta_1,\eta_2,r_1,r_2,\cos\theta).
\end{equation}
The function in the right-hand side of the above equation can be
expanded in general  as
\begin{eqnarray}
  F(\eta_1,\eta_2,r_1,r_2,\cos\theta)&=&\sum_l {2l+1\over 4\pi} 
  C(\eta_1,\eta_2,r_1,r_2,l) P_l(\cos\theta)
\nonumber
\\
  &=&\sum_{l}\sum_{m=-l}^{l}
  C(\eta_1,\eta_2,r_1,r_2,l) Y^*_{lm}(\Omega_{\bft_1})
       Y_{lm}(\Omega_{\bft_2}) .
\label{Ap2}
\end{eqnarray}
Substituting the above expansion into equation (\ref{appB2})
and integrating over $d\Omega_{\bft_1}$ and $d\Omega_{\bft_2}$, 
we obtain equation (\ref{eq:q12}):
\begin{equation} 
  \bigl<\Delta_{k_1l_1m_1}(\eta_1) 
  \Delta_{k_2l_2m_2}^*(\eta_2)\bigr> 
  = Q(k_1,k_2,\eta_1,\eta_2,l_1)
  \delta_{l_1l_2}\delta_{m_1m_2},
\end{equation}
where 
\begin{equation} 
Q(k_1,k_2,\eta_1,\eta_2,l) = 
\int  r_1^2 dr_1 \int  r_2^2 dr_2 
C(\eta_1,\eta_2,r_1,r_2,l) X_{k_1 l}(r_1) X_{k_2 l}(r_2) .
\end{equation}

\end{appendix}

\clearpage

\clearpage

\clearpage

\clearpage

\clearpage

\end{document}